\documentclass[aps,prl,twocolumn,groupedaddress]{revtex4-1}

\usepackage{graphicx}
\usepackage{amssymb}
\usepackage{amsmath}

\begin{document}

\title{Charge Ordering Geometries in Uniaxially-Strained NbSe$_2$}

\author{Felix Flicker}
  \email{flicker@physics.org}
  \affiliation{University of Bristol, H. H. Wills Physics Laboratory, Tyndall Avenue, Bristol, BS8 1TL, UK}
\author{Jasper van Wezel}
  \email{vanwezel@uva.nl}
  \affiliation{Institute for Theoretical Physics, University of Amsterdam, 1090 GL Amsterdam, The Netherlands}

\date{\today}

\begin{abstract}
Recent STM experiments reveal niobium diselenide to support domains of striped (1Q) charge order side-by-side with its better-known triangular (3Q) phase, suggesting that small variations in local strain may induce a quantum phase transition between the two.
 We use a theoretical model of the charge order in NbSe$_2$, based on a strong momentum- and orbital-dependent electron-phonon coupling, to study the effect of uniaxial strain. We find that as little as $0.1\%$ anisotropic shift in phonon energies breaks the threefold symmetry in favour of a 1Q state, in agreement with the experimental results. The altered symmetries change the transition into the ordered state from weakly-first-order in the 3Q case, to second order in the 1Q regime. Modeling the pseudogap phase of NbSe$_2$ as the range of temperatures above the onset of long-range order in which phase coherence is destroyed by local phonon fluctuations, we find a shortening of the local ordering wavevector with increasing temperature, complementing recent X-ray diffraction observations within the low-temperature phase. 
\end{abstract}

\maketitle

The stability of different charge density wave (CDW) geometries is a topic of much current interest. In the simplest case a `1Q' CDW selects a single preferential direction~\cite{Gruner88}. In two-dimensional materials, two perpendicular 1Q CDWs can coexist, creating a 2Q  pattern. In the layered high-temperature superconductors such a checkerboard state is believed to compete with 1Q order~\cite{KivelsonEA03,MelikyanNorman,RobertsonEA06,SeoEA07}; recently the 1Q state was found to dominate the 2Q in YBCO~\cite{CominEA15}. The layered hexagonal structure of niobium diselenide ($2H$-NbSe$_2$) allows instead for a 3Q triangular CDW with three superposed 1Q patterns~\cite{WilsonEA74,RossnagelEA01,WeberEA11,WeberEA13}. Recent surface studies via scanning tunnelling microscopy (STM) suggest a quantum phase transition between 3Q and 1Q states in NbSe$_2$ may be tuned by local strain, offering the possibility of studying the interplay between the two, as well as a direct comparison to the high-T$_c$ case~\cite{ArguelloEA14,SoumyanarayananEA13}.  

It was recently shown that the minimal model reproducing the full range of experimental observations on the 3Q ordered state in NbSe$_2$ includes a strong electron-phonon coupling, dependent on both the momenta and the orbital characters of the electronic states scattered between~\cite{us}. Motivated by the recent STM observations, in this Letter we employ the same model in a free energy analysis to study the charge-ordered state upon application of uniaxial strain~\cite{SoumyanarayananEA13}. We find that while the 3Q state, which respects the threefold lattice symmetry, is the lowest-energy configuration in the unstrained case, little uniaxial strain is required to break the symmetry to 1Q, agreeing with suggestions that lattice defects may locally stabilize the 1Q state~\cite{SoumyanarayananEA13}. We find that the 1Q phase transitions are second-order, whereas the 3Q transitions are weakly-first-order. With increasing temperature we predict a shortening CDW wavevector in the locally-fluctuating, short-range-ordered, pseudogap regime above the CDW transition, analogous to the recently-observed non-monotonic evolution of the CDW wavevector within the ordered state~\cite{FengEA15}.

\begin{figure}  
  \includegraphics[width=\columnwidth]{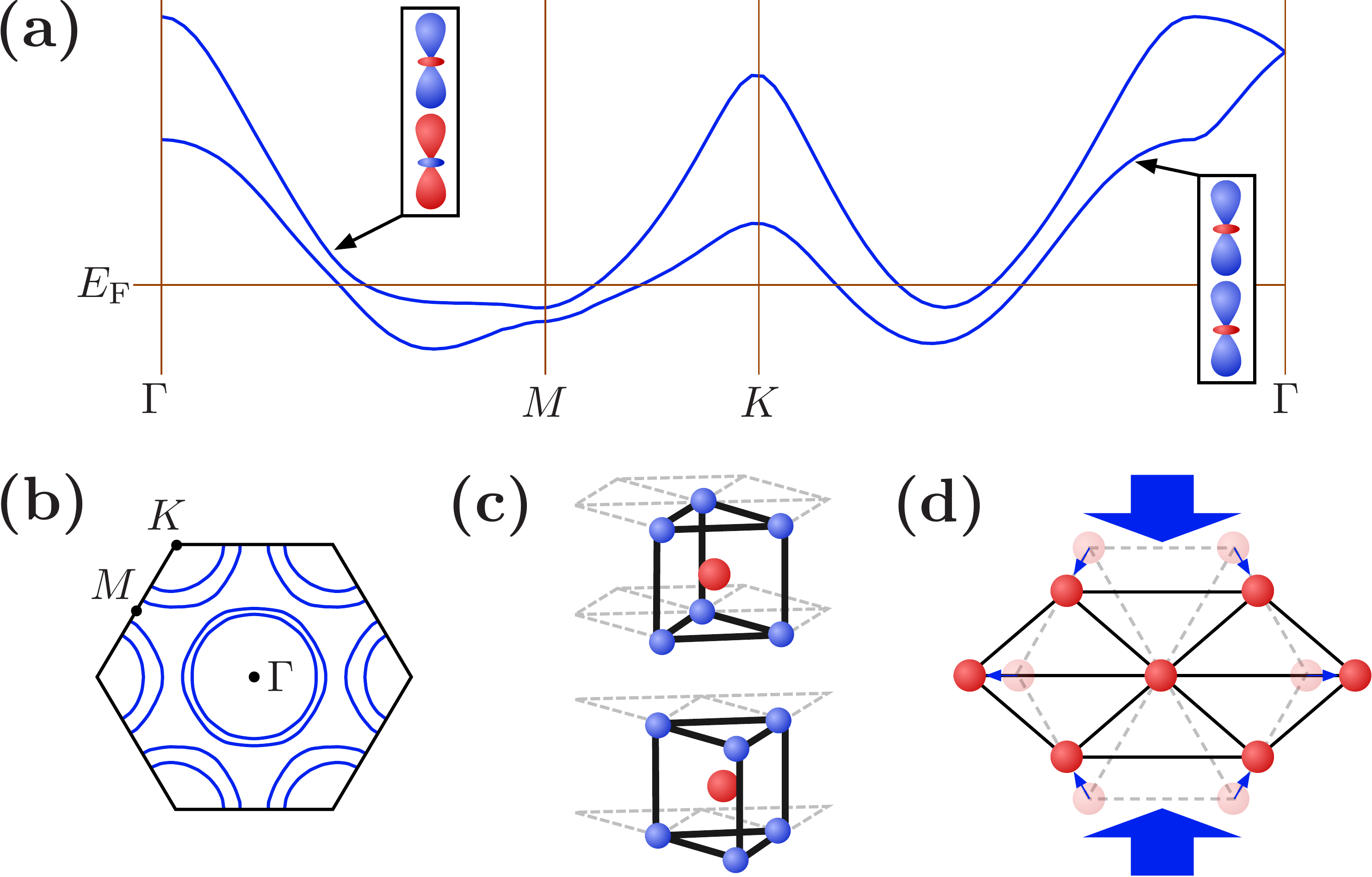}  
  \caption{\label{bandstructure} 
  {\bf (a)} The bandstructure of NbSe$_2$ modeled by the tight-binding fit described in Ref.~\onlinecite{us}. Only the bands crossing $E_{\text{F}}$ are shown, corresponding primarily to bonding and anti-bonding combinations of the two Niobium $d_{3z^2-r^2}$ orbitals within a unit cell. Only the lower band is involved in the CDW formation. {\bf (b)} The Fermi surface consists of concentric barrel-shaped pockets around the $\Gamma$ and $K$ points~\cite{us,DoranEA78,Doran78,RossnagelEA01}. We ignore a small pocket around $\Gamma$. {\bf (c)} The layered structure of NbSe$_2$, with Nb in red and Se in blue. {\bf (d)} The atomic displacements of the crystal structure caused by uniaxial strain within the plane of Nb atoms, indicated by the large blue arrows. Keeping the volume of the unit cell constant and increasing the horizontal interatomic distances to $x_0+\epsilon$ causes the diagonal interatomic distances to become approximately $x_0-\epsilon/2$.
  }
  \end{figure}

\emph{Model}--niobium diselenide consists of hexagonal layers of niobium atoms sandwiched between layers of selenium atoms, displaced so that they lie above and below half of the Nb interstitial locations. The crystal structure and bandstructure are depicted in Fig.~\ref{bandstructure}. Consecutive sandwich layers are displaced to have the complementary half of the interstices occupied, giving two formula units per unit cell. 3Q charge order develops below $T_{\textrm{CDW}}=33.5\,$K. Recent STM measurements show stable domains of 1Q order on the surface of NbSe$_2$, coexisting with neighboring regions of 3Q order, and occurring in areas where the topmost niobium layer is slightly raised~\cite{SoumyanarayananEA13}. This suggests that the 1Q state may be stabilized by local strain, causing a quantum phase transition between the 1Q- and 3Q-ordered states~\cite{SoumyanarayananEA13}. The CDW wavevector $\mathbf{Q}_{\textrm{CDW}} = \left(1-\delta\right)\frac{2}{3}{\Gamma M}$ is found to be incommensurate with the lattice in both geometries, with $\delta_{3Q}\approx0.014$ and $\delta_{1Q}\approx0.143$~\cite{WilsonEA74,SoumyanarayananEA13}.

It has previously been shown that the CDW transition in NbSe$_2$ can be successfully described by a strong coupling between electrons $\psi$ and atomic displacements $\varphi$ through the interaction~\cite{us}:
\begin{align}\label{Hint}
\hat{H}_{\text{int}}=\sum_{\mu\nu\mathbf{kq}}\mathbf{g}^{\mu,\nu}_{\mathbf{k},\mathbf{k}+\mathbf{q}}\hat{\varphi}^{\dagger}_{\mathbf{q}}\hat{\psi}_{\mathbf{k}}^{\nu \dagger} \hat{ \psi }_{ \mathbf { k } +\mathbf{q}}^{\mu \phantom{\dagger}},
\end{align}
where the electron-phonon coupling $\mathbf{g}^{\mu,\nu}_{\mathbf{k},\mathbf{k}+\mathbf{q}}$ depends on the momenta of both the ingoing and outgoing electron states and the bands $\mu,\nu$ scattered between. The analytic expression for $\mathbf{g}$ has been shown to be applicable to a range of transition metal compounds including NbSe$_2$, and accounts for the orbital characters of the electron bands with dispersion $\xi^\mu_{\mathbf{k}}$ crossing the Fermi level~\cite{VarmaEA79,us}. It yields a stronger electron-phonon coupling in the lower band than the upper, explaining the order-of-magnitude difference in CDW gap size of the two bands in angle-resolved photoemission spectroscopy (ARPES)~\cite{BorisenkoEA09,RahnEA12,us}. We restrict attention to the lowest band from here on.
This model reproduces the full host of available experimental observations on the 3Q state, with the overall strength of the electron-phonon coupling as the only free parameter (chosen to give the observed transition temperature $T_{\textrm{CDW}}=33.5\,$K)~\cite{us}. 

Following reference~\cite{WeberEA13} we model the (isotropic) bare phonon dispersion $\Omega_0\left(|\mathbf{q}|\right)$ by a Brillouin function with a maximum of $11.2\,\textrm{meV}$ at the zone boundary. Since the longitudinal phonon is seen to soften in inelastic x-ray scattering~\cite{WeberEA13}, we restrict attention to $g=\hat{\mathbf{q}}\cdot\mathbf{g}$. Uniaxial strain is modeled as an anisotropic linear shift in bare phonon frequencies. As shown in Fig.~\ref{bandstructure}(d), stretching the lattice in one direction causes a contraction in the perpendicular direction. We correspondingly define a strain parameter $\sigma$ through $\Omega_1=\Omega_0\left(1+\sigma\right)$ along one of the $\mathbf{Q}_{CDW}\parallel{\Gamma M}$ directions, and $\Omega_{2,3}=\Omega_0\left(1-\frac{\sigma}{2}\right)$ in the $2\pi/3$-rotated directions.

\emph{Landau Theory}--the free energy can be expanded in terms of the three order parameters $\varphi_j=\langle\hat{\varphi}\left(\mathbf{q}_j\right)\rangle$, where $\mathbf{q}_j=\left(1-\delta\right)\frac{2}{3}{\Gamma M}_j$, and $j$ labels the three inequivalent ${\Gamma M}$ directions. At any given value for $\delta$, the most general free energy to fourth order in $\varphi_j$ is:
\begin{align}\label{F}
\beta F\left[\varphi_1,\varphi_2,\varphi_3\right] =& -\sum_{j=1}^3\left\{ a\left(\Omega_{j}\right) \left|\varphi_{j}\right|^2 \right\} - b\varphi^{\phantom{2}}_{1} \varphi^{\phantom{2}}_ {2} \varphi^{\phantom{2}}_{3} \notag \\
& + c \sum_{j\textrm{mod}3}\left\{ \left|\varphi_{j}\right|^{4} + d |\varphi_{j}|^{2}|\varphi_{j+1}|^{2} \right\}.
\end{align}
The phases of $\varphi_j$ are chosen such that the third-order term is real and $b$ is positive. Starting from the Hamiltonian of Eq.~\eqref{Hint}, the coefficients $a$-$d$, and their dependences on $\mathbf{q}$ and $\beta=1/T$, can be evaluated numerically~\cite{MelikyanNorman}. Strain enters via the $\Omega_j$ dependence of the quadratic coefficient.

In the 3Q phase the order parameters obey $\varphi_1=\varphi_2=\varphi_3 \equiv \varphi$, while in each of the 1Q phases only a single $\varphi_j$ is nonzero. The free energy then takes the forms:
\begin{align}\label{F1Q3Q}
\beta F^j_{\textrm{1Q}}\left[\varphi_j\right] = &-a\left(\mathbf{q}_j,T,\Omega_j\right)|\varphi_j|^{2} +c\left(\mathbf{q}_j,T\right)|\varphi_j|^{4} \notag \\
\beta F_{\textrm{3Q}}\left[\varphi\right] = &-\sum_{j=1}^{3} a\left(\mathbf{q}_j,T,\Omega_j\right)|\varphi|^{2} -b\left(\mathbf{q},T\right)\varphi^{3}\notag\\
&+3 c \left(\mathbf{q},T\right)\left(1+d\left(\mathbf{q},T\right)\right)|\varphi|^{4},
\end{align}
where $b$-$d$ do not depend on strain, and can be evaluated along any of the ${\Gamma M}$ directions. Boundedness of the free energy requires $c>0$, which is the case in all calculations described below.

Starting from a disordered state with $a<0$, minimization of Eq.~\eqref{F1Q3Q} yields a second-order phase transition into either a 3Q or a 1Q state whenever $a$ becomes positive. If $a$ stays negative, but $ c\left(1+d\right)<-\frac{b^2}{18a}$, a 3Q state forms via a first-order transition. Both scenarios require $d>-1$, which is the case for all calculations considered. For any nonzero $b$, approaching a second-order transition into the 3Q state at $a=0$ causes a divergence in ${b^2}/{a}$, which forces a first-order transition into the 3Q state instead. The two (would-be) transitions are typically close in temperature and strain, suggesting a weakly-first-order transition~\cite{LarkinPikin}. We therefore only expect either a second-order transition from disorder into a 1Q state, or a first- or weakly-first-order transition from disorder into the 3Q state. 

\emph{Temperature Dependence}--the quadratic coefficient in Eq.~\eqref{F1Q3Q} describes the electronic susceptibility towards the formation of charge order, given by:
\begin{align}\label{D2}
a \left(\mathbf{q}_j,T,\Omega_{j}\right) &= -\Omega_{j}\left(\mathbf{q}_j\right)+\chi \left(\mathbf{q}_j,T\right), \\
\chi\left(\mathbf{q}_j,T\right) &= -\sum_{\mathbf{k}}\left|g_{\mathbf{k},\mathbf{k}-\mathbf{q}}\right|^{2}\mathfrak{Re}\left[\frac{f\left(\xi_{\mathbf{k}}
\right)-f\left(\xi_{\mathbf{k}-\mathbf{q}}\right)}{\xi_{\mathbf{k}}-\xi_{\mathbf{k}-\mathbf{q}}-i\delta}\right]. \notag
\end{align}
where $f\left(\xi_\mathbf{k}\right)$ is the Fermi-Dirac distribution, and $\delta\rightarrow0^+$. The definition of the `generalized susceptibility' $\chi$ follows Refs.~\onlinecite{VarmaWeber77,Doran78,us}, and is expressed here within the Random Phase Approximation (RPA). The usual (bare electronic) susceptibility, $\chi_0\left(\mathbf{q},T\right)$, is given by $\chi$ with $g$ set to unity. Note that $\sum_j a \left(\mathbf{q}_j,T,\Omega_{j}\right)$, appearing in $F_{\textrm{3Q}}$, is independent of strain. CDW order first sets in when $a$ changes sign, meaning $\chi$ has renormalized the phonon to zero energy at a nonzero wavevector $\mathbf{q}=\mathbf{Q}_{\textrm{CDW}}$. At zero strain, $\mathbf{Q}_{\textrm{CDW}}$ within RPA agrees with the experimentally-observed CDW wavevector~\cite{us}. 

\begin{figure}  
  \includegraphics[width=0.98\columnwidth]{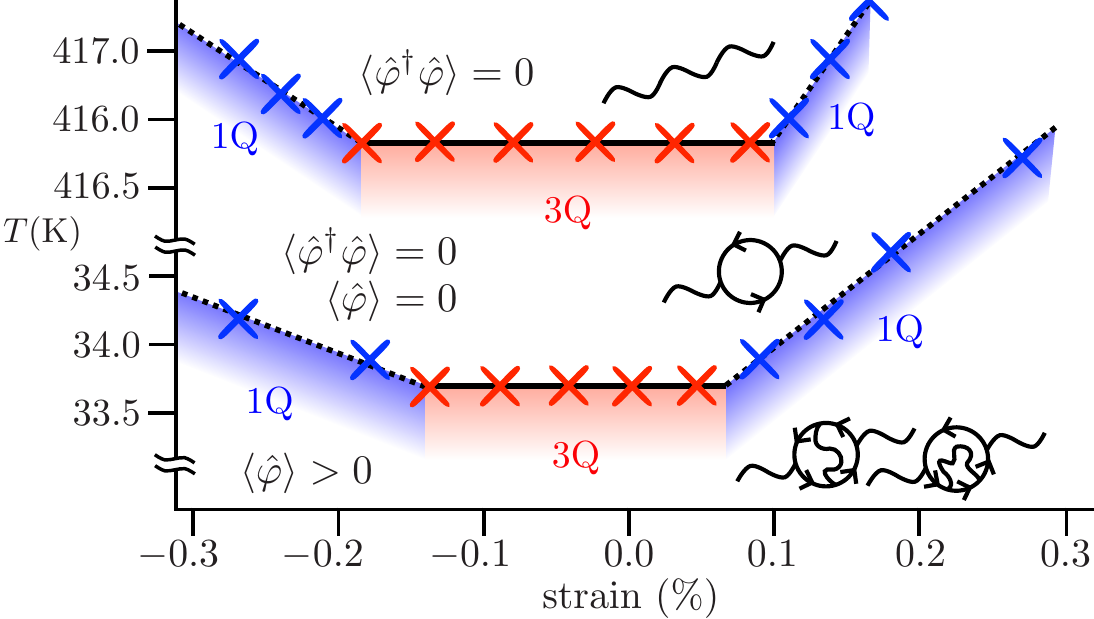}  
    \caption{\label{phaseDiagram}
The phase diagram of NbSe$_2$ as a function of uniaxial strain and temperature. Cooling from high temperature, the disordered system develops fluctuating short-ranged order when the RPA predicts a transition, but long-range phase coherence is suppressed by phonon fluctuations in the MMA. As the system cools further, long-ranged CDW order sets in at $T_{\textrm{MMA}}$. The order changes from 3Q to 1Q at an anisotropy in phonon energy of about $0.1\%$. Dashed black lines indicate second-order phase transitions, whereas solid black lines indicate weakly-first-order transitions.
  }
  \end{figure}

To properly account for the system's temperature dependence, it is necessary to include the entropy of the phonon field~\cite{McMillan1977_2}, by introducing internal phonon lines into the RPA diagrams~\cite{Inglesfield80,YoshiyamaEA86,VarmaSimons83}. To fourth order in the coupling $g$, two such terms contribute to $\chi$, shown in Fig.~\ref{phaseDiagram}. Their self-consistent inclusion forms the Mode-Mode Coupling Approximation (MMA), known to give more accurate temperature dependences than RPA in a range of transition metal compounds~\cite{Inglesfield80,YoshiyamaEA86,VarmaSimons83}. We neglect one of the diagrams by appeal to Migdal's theorem, since it contains only a vertex correction. The remaining diagram, constituting a self-energy correction to the electrons, is responsible for the observed suppression of the RPA transition temperature in NbSe$_2$~\cite{us}.

The range of temperatures in which RPA predicts order, while MMA predicts the order to be suppressed by the entropy of the phonon field, can be interpreted as a pseudogap regime~\cite{us}. In this regime the local amplitude of the displacements, $\sqrt{\langle\hat{\varphi}^{\dagger}\hat{\varphi}\rangle}$, is nonzero, while the order parameter itself, $\langle\hat{\varphi}\rangle$, averages to zero through fluctuations in phase~\cite{YoshiyamaEA86,McMillan75,VarmaSimons83}. The result is fluctuating short range order without long-range phase coherence~\cite{Chatterjee2015}. 

\emph{The CDW Wavevector}--numerically evaluating the coefficients in Eq.~\eqref{F1Q3Q}, within RPA, yields a weakly-first-order phase transition from a high-temperature disordered state to a 3Q CDW at $T_{\textrm{RPA}}=416\,$K. Including phonon fluctuations via MMA suppresses the order, resulting in fluctuating regions of 3Q order without long-range coherence. At $T_{\textrm{MMA}}=33.5\,$K we find a weakly-first-order transition in MMA, again to a 3Q state. This is interpreted as the onset of long-range phase-coherence~\cite{Chatterjee2015}, as indicated in Fig.~\ref{phaseDiagram}.
  
The CDW wavevector $\mathbf{Q}_{\textrm{CDW}}$ coincides with the maximum of the quadratic coefficient $a\left(\mathbf{q}_j\right)$ in the free energy, or equivalently of the generalized susceptibility $\chi$ of Eq.~\eqref{D2}. The shape of $\chi$ is a convolution of the squared electron-phonon coupling with the bare electronic susceptibility $\chi_0$. The CDW wavevector is selected either by a divergence in $\chi_0$ due to Fermi surface nesting, by a strongly-peaked dependence of the electron-phonon coupling on electron momenta, or by a combination of the two; in NbSe$_2$, $\mathbf{Q}_{\textrm{CDW}}$ is selected predominantly by the wavevector dependence of $g$~\cite{us,WeberEA11}.
  
\begin{figure}  
  \includegraphics[width=0.98\columnwidth]{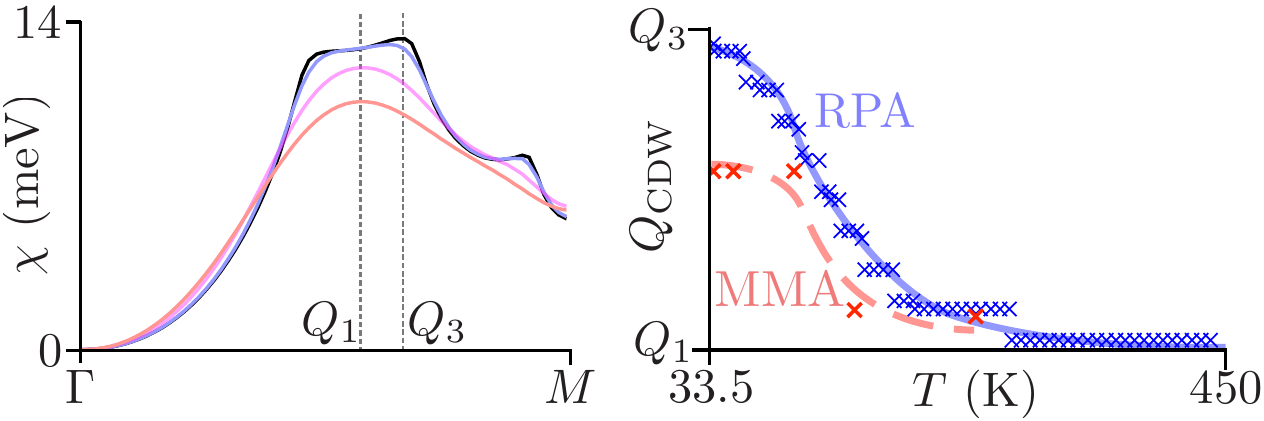}  
  \caption{\label{figQ}
Left: the generalized susceptibility $\chi$ along $\Gamma M$ in the RPA at zero strain. From top to bottom, the black curve is at $33.5\,$K, blue $100\,$K, purple $303.5\,$K, and red $500\,$K. Right: the peak position of $\chi$, which dictates the CDW ordering vector within the fluctuating, short-range-ordered regions, decreases with increasing temperature. Both the results of the MMA and RPA calculations are indicated. The solid and dashed lines are guides to the eye only.
  }
  \end{figure}

While the MMA free energy expansion is strictly valid only close to the transition temperature, the susceptibility can still be evaluated within the same approximation at temperatures above $T_{\textrm{MMA}}$. Similarly the RPA expansion is valid only slightly above $T_{\textrm{RPA}}$, but the susceptibility can still be evaluated within the RPA-ordered regime. The peaks of $\chi$ within these approximations provide an estimate for the CDW wavevector within the locally-fluctuating short-range-ordered region characterizing the pseudogap regime. As temperature increases, $\chi_0\left(\mathbf{q}_j\right)$ broadens, and peaks in the temperature-independent electron-phonon coupling begin to dominate $\chi$. The result is a decrease in the local $\left|\mathbf{Q}_{\textrm{CDW}}\right|$ as temperature increases in the range $T_{\textrm{MMA}}<T_{\textrm{CDW}}<T_{\textrm{RPA}}$, shown in Fig.~\ref{figQ}. A crossover between two distinct regimes, one close to $T_{\textrm{MMA}}$ and the other at high temperatures, can clearly be seen. In the former regime, the bare susceptibility can modify the detailed momentum dependence of the electron-phonon coupling and cause small local extrema to appear in the full susceptibility, thus setting the precise value of $\mathbf{Q}_{\textrm{CDW}}$ at low temperatures. Beyond the crossover, temperature has smoothed these local extrema, and the CDW wavevector is determined solely by the remaining momentum dependence of the electron-phonon coupling. This evolution complements recent X-ray diffraction observations within the ordered phase, of a non-monotonic evolution of $\mathbf{Q}_{\textrm{CDW}}$ as temperature increases towards $T_{\textrm{CDW}}$ from below~\cite{FengEA15}. 

\emph{Phase Diagram}--tracking the coefficients of the free energy as functions of uniaxial strain and temperature yields the phase diagram of Fig.~\ref{phaseDiagram}. The range of temperatures between $T_{\textrm{RPA}}$ and $T_{\textrm{MMA}}$ again corresponds to a pseudogap regime characterized by fluctuating local order~\cite{Chatterjee2015,us}. Starting from unstrained conditions, it takes of order $0.1\%$ anisotropic change in the bare phonon energy to break the 3Q symmetry to 1Q. To estimate the corresponding lattice deformations, we use the in-plane Gr\"{u}neisen parameter, $-\left(a/\Omega_0\right)\textrm{d}\Omega_0/\textrm{d}a$, with in-plane lattice parameter $a$. For several layered hexagonal materials related to NbSe$_2$, including MoS$_2$, boron nitride, and graphite, the Gr\"{u}neisen parameter is of order unity~\cite{ConleyEA13,MounetEA05}. This suggests that a $0.1\%$ anisotropic phonon energy shift also corresponds to changes of order $0.1\%$ in the inter-atomic distances. 

The transition temperature into the 1Q phase increases with uniaxial strain, as the breaking of three-fold rotational symmetry stabilizes 1Q order. The rates of increase at positive and negative strain differ because, as indicated in Fig.~\ref{bandstructure}(d), the atomic displacements stabilizing the 1Q phase are proportional to $\sigma$ at positive strain, when the CDW order develops along the strain direction, but proportional to $-\sigma/2$ for negative strain, with CDW order at an angle of $2\pi/3$ to the direction of applied strain. We found the 2Q state, with two CDWs at $\pm 2\pi/3$ to the strain direction, to have higher free energy than either the 1Q or the 3Q state within the range of parameters relevant to NbSe$_2$~\cite{MonctonEA77}.

At $T_{\textrm{CDW}}=T_{\textrm{MMA}}$ and zero strain, the system orders into a 3Q geometry via a weakly-first-order transition. Within the MMA, phonon fluctuations renormalize the generalized susceptibility, yielding a decreased range of stability for the 3Q state as compared to the RPA. Application of just under $0.1\%$ strain then stabilizes the 1Q CDW, in agreement with STM findings of a 1Q phase in regions with an upper-bound uniaxial strain of about $0.45\%$~\cite{SoumyanarayananEA13}. We predict this 1Q order to also be stabilized in the bulk upon application of uniaxial pressure to the entire sample.
 
\emph{Conclusions}--We compared the stability of proximate 1Q- and 3Q-ordered CDW phases  in niobium diselenide using a Landau free energy expansion, based on a tight-binding model that was previously shown to capture all of the various known properties of the zero-strain, 3Q-ordered phase~\cite{us}. By comparing the model's phase diagram within the RPA and MMA, it is found that phonon fluctuations, present in the latter, suppress long-range order and lead to the formation of a pseudogap regime between $T_{\textrm{RPA}}$ and $T_{\textrm{MMA}}$ in which the order parameter has a nonzero local magnitude but no long-range phase coherence. Both the pseudogap and CDW phases are shown to have a 3Q geometry, but it takes only of order $0.1\%$ uniaxial strain to induce a 1Q geometry, in agreement with suggestions that the 1Q state seen in STM is stabilized by surface-layer distortions causing local strain~\cite{SoumyanarayananEA13}. We propose that the bulk quantum phase transition between the different ordering geometries can be accessed by the application of comparable amounts of bulk uniaxial strain.

The free energy expression shows that while, in principle, the 3Q phase transition can be either first- or second-order, any would-be second-order transition is pre-empted by a weakly-first-order transition. We found that the 3Q phase at low strain is of this weakly-first-order form, in agreement with previous experimental and theoretical studies~\cite{McMillan75,McMillan76,McMillan77,MonctonEA75,MonctonEA77,HarperEA75,Rice80}, while the transition from disorder to the strained 1Q phase is second order.

The CDW wavevector $\mathbf{Q}_{\textrm{CDW}}$ is dictated by the location of the maximum in the momentum-dependent generalized susceptibility $\chi$, implying that within the short-range-ordered domains of the pseudogap, $\mathbf{Q}_{\textrm{CDW}}$ can be tracked by the peak in $\chi$ between $T_{\textrm{MMA}}$ and $T_{\textrm{RPA}}$. As temperature increases within the pseudogap, the momentum dependence of the electron-phonon coupling increasingly dominates the bare electronic susceptibility in determining the shape of $\chi$, resulting in a decrease of the CDW wavevector within the locally-fluctuating regions. This complements recent experimental work in which a non-monotonic evolution of the CDW wavevector was found within the ordered state itself~\cite{FengEA15}. 

The situation in NbSe$_2$ recalls that in the layered high-temperature superconductors, which exhibit both charge order~\cite{EmeryEA99,WalstedtEA11,ChangEA12,GhiringhelliEA12} -- believed to be close to the phase boundary between the 1Q and 2Q geometries supported by their square lattices~\cite{MelikyanNorman,CominEA15} -- and pseudogaps characterized by some form of fluctuating local order~\cite{KivelsonEA03,MelikyanNorman,RobertsonEA06,SeoEA07,Mesaros2011}. Contrary to the high-T$_c$ case, the underlying physics giving rise to these features in NbSe$_2$ is well understood~\cite{us}. The interplay of the CDW geometries with external parameters such as applied uniaxial stress may now be used as a test case from which to consider the geometries of charge-ordered states more generally, and may thus shed some light on the case of the high-T$_c$ superconductors.

\emph{Acknowledgments}--the authors wish to thank M. R. Norman and B. A. C. Amorim for many useful discussions. JvW acknowledges support from a 
VIDI grant financed by the Netherlands Organisation for Scientific Research (NWO).

\bibliographystyle{apsrev4-1}
\bibliography{mybib}

\end{document}